# A comprehensive view on the manifestations of aggregate demand and aggregate supply shocks in Greece, Ireland, Italy and Portugal

**Ionuţ JIANU**
Bucharest University of Economic Studies, Romania
ionutjianu91@yahoo.com

**Abstract.** *The main goal of the paper is to extract the aggregate demand and aggregate supply shocks in Greece, Ireland, Italy and Portugal, as well as to examine the correlation among the two types of shocks. The decomposition of the shocks was achieved by using a structural vector autoregression that analyses the relationship between the evolution of the gross domestic product and inflation in the period 1997-2015. The goal of the paper is to confirm the aggregate demand - aggregate supply model in the above-mentioned economies.*

## 1. Introduction

Adhesion to the European Union (EU) involves the obligation of the member countries to adopt the Euro currency according to a target date set by them (except for the United Kingdom and Denmark, as per the opt-out clauses applied to them), following the fulfillment of the nominal convergence criteria and joining the currency exchange mechanism, which represents the "purgatory" of the European community space, a mechanism in which the member countries must prove their stability for a two-year period.

When the Euro currency is adopted, the members of the Euro zone make a sovereignty transfer to the European Central Bank (ECB), through the common monetary policy, which determines the Economic and Monetary Union (EMU) interest rate. For that reason, the correlation of the business cycles becomes a very important variable, as ECB shall take the measures favorable to the Euro zone. To be more precise, if a state is in a recessionary gap, and the Euro zone is in an expansionist phase of the business cycle, the applied monetary measure will consist in increasing the interest rate, which further deepens the negative phase of the output gap manifesting in the divergent state.

An optimal monetary zone also involves synchronizing the structural shocks between the member states. The asymmetry of aggregate demand and supply shocks invalidates from the very beginning the assumption of an optimal monetary zone in the case of the EMU. The shocks of demand, depending on their manifestation period on the territory of an economy, may be temporary or permanent.

Aggregate demand comprises the aggregate of the production components, calculated according to the expenditure approach, such as private consumption, government expenditures, investments and net exports, while aggregate supply is the result of adding the value of imports to the value of the gross domestic product (GDP). According to the economic theories, a positive demand shock influences the GDP in the same direction, but the effect is temporary. The impact on prices is also favorable. On the other hand, a positive supply shock results in price decline and in production reaching another positive balance level.

The reason and timeliness of this topic lies in the important role played by the real convergence in determining the winners and losers of the Economic and Monetary Union. The analysis focuses on the peripheral group of countries comprising Portugal, Ireland, Italy and Greece, which had to face structural shortcomings and fully experienced the shock of the public debt crisis. This group of countries shall be referred to as PIIG.

The general goal of the paper is to review the effects of aggregate demand and supply shocks on the GDP or inflation levels in the analyzed countries, by using a structural vector autoregression (SVAR). This will be achieved by undertaking the following specific goals:

a) determining the optimal number of lags, following the estimation of the vector autoregression (VAR), including the endogenous variables related to GDP and inflation, for each analyzed country;
b) identifying the economic restrictions and running the structural factorization;
c) verifying the aggregate demand - aggregate supply model based on the impulse - response function;
d) determining the correlation coefficients for aggregate demand and aggregate supply shocks.

This analysis aims at investigating the way in which the aggregate demand - aggregate supply model applies to PIIG countries. Creating a common destiny involves the manifestation of correlated shocks for aggregate demand (similar economic policies) and for aggregate supply (the synchronism of the technological shocks or those that influence productivity).

## 2. Literature review

Blanchard and Quah (1988) estimated a SVAR composed of the differential of the logarithm of GDP and the logarithm of unemployment rate to decompose structural innovations. They identified the long-run restriction, whereby a demand shock influences the production level temporarily and not permanently, with the effect being neutralized after a certain number of quarters. The authors found that the application of the long-run restriction and the restriction related to the normalization of shocks is sufficient to accurately determine the SVAR. Following the correct identification of the restrictions, Blanchard and Quah decomposed the shocks of demand and supply according to the relation *Ae = Bu*.

Bayoumi and Eichengreen (1992) considered the differential of the logarithm of GDP and that of the logarithm of GDP deflator, observing the previously mentioned methodology and found that the economy of the United States of America (USA) adjusts to shocks quicker than the economy of the European Union, while structural shocks were also more correlated in the US economy. In other words, it is more difficult for the EMU economy to operate as an optimum currency area than for the USA. They also identified discrepancies between the supply shocks affecting Germany, France, Belgium, Netherlands and Denmark, and those in the PIIGS countries. Bayoumi and Eichengreen furthered the concept of asymmetrical shock by examining the neutralizing capacity of the shocks.

According to Cooley and Dwyer (1998), the SVAR model must meet four conditions to be correctly represented, as follows: (I) the presence of a stationary variable and of a non-stationary variable at the initial level, (II) the assumption that structural innovations are not inter-correlated, (III) a VAR with p lags must represent the data dynamics and (IV) the assumption pertaining to the temporary impact of a structural shock on a variable included in the model.

The correlation of supply shocks is much more important than demand shocks correlation, regarding the construction of an optimum currency area, according to Fidrmuc and Korhonen (2003), but their results identified asymmetrical correlations between the European Union countries. They identified a higher synchronization of the structural shocks manifesting in Italy and Portugal than those in the Euro zone.

Frenkel and Nickel (2002) determined the correlation degree of structural shocks, with the demand shocks correlation ranking higher than the synchronism of supply shocks in the country pairs Italy-Greece and Italy-Spain in the period 1995-2001.

Following the methodology of Blanchard and Quah, Bergman (2005) analyzed the decomposition method of structural shocks following the application of long-run restriction. The economist proved that the impulse-response function can use incorrect signs that are incompatible with the theoretical model. He also refers to the possible existence of multiple structural shocks, which can provide unwanted empirical results.

Dinu and Marinaș (2006), as well as Socol et al. (2010), analyzed the southern EU model and found the lowest efficiency level of the income redistribution systems therein, as the peripheral countries encouraged long-term unemployment by their economic structures.

According to the study developed by Marinaș (2012), the volatility of the demand shocks was higher than the volatility of the supply shocks until 2009, and vice versa afterwards. After reviewing the impulse-response function, he found that the neutralization of the supply shocks on economic growth occurs in approximately five years in the case of Denmark, Ireland, Spain and Portugal. Marinaș also identified a higher demand shocks correlation for the country pair Greece-Ireland.

## 3. Methodology

In order to decompose the aggregate demand and supply shocks, I used the methodology proposed by Blanchard and Quah (1989) and subsequently furthered by Bayoumi and Einchengreen (1992), who decomposed the shocks influencing production and inflation into shocks of aggregate demand and aggregate supply, through a SVAR and an impulse-response function.

For such purpose, I extracted the quarterly statistical data published by Eurostat related to volume index of the gross domestic product (index 2010=100) and GDP deflator (index 2010=100) for PIIG countries. The extracted data series cover the period 1997-2015, as the goal was to analyze the Pearson statistical correlation by sub-periods (1997-2006 and 2007-2015), a method aimed at capturing the impact of certain significant events in those economies, such as the onset of the global financial crisis. Even if the apex of the economic crisis was felt in the European Union in the first quarter of 2009, the effects started to emerge in late 2007. The sub-periods were chosen so to maintain a balance in terms of their duration.

This research includes a total of 76 observations, which were processed in Eviews 7.0 software. I used Tramo-Seats, in order to eliminate the influence of seasonality for both GDP and GDP deflator series, subsequently, performing the logarithm of the resulting variables, this operation having the role to reflect the growth rate.

A valid representation of an SVAR first involves the transformation of the data series into stationary variables. Augmented Dickey-Fuller test was used to verify the order of integration for both variables, and subsequently, I performed the appropriate transformations, by differentiation to ensure the stationarity of the variables.

In this case, a bivariate SVAR may be represented as follows:

$$y_t = \mu_{10} - \rho_{12}\pi_t + \rho_{11}y_{t-1} + \rho_{12}\pi_{t-1} + u_{y_t} \tag{1}$$

$$\pi_t = \mu_{20} - \rho_{21}y_t + \rho_{21}y_{t-1} + \rho_{22}\pi_{t-1} + u_{\pi_t}$$

where: $y_t$ represents the differential of the logarithmized series of the gross domestic product, $\pi_t$ reflects the differential of the logarithm of the GDP deflator, $y_{t-1}$ and $\pi_{t-1}$ are the previously mentioned series, lagged by one quarter, $u_{y_t}$ and $u_{\pi_t}$ outlines the structural innovations of variables, while $\mu$ and $\rho$ represent the model coefficients.

I created a VAR model for each country included in the analysis, using as endogenous variables the GDP indicator (measured in volume) and GDP deflator, resulting 4 separate models. As for choosing the lag, I applied Lag Length Criteria function for selecting the appropriate lag, using a limit of 8 lags, since the study was conducted using quarterly data. In exceptional situations, I used Lag Exclusion Wald Test for the selected lag. Depending on the preferential lag assigned by Likelihood Ration, Final prediction error, Akaike information criterion, Schwarz information criterion and Hannan-Quinn information criterion tests, I have determined the optimal lag.

A SVAR is a research tool appropriate to identifying the structural shocks and applying restrictions according to the economic theory. Following the methodology of Blanchard and Quah, the vector composed by the two variables related to the differential of the logarithm of GDP and GDP deflator can be written according to aggregate demand shocks ($u_{dt}$) and aggregate supply shocks ($u_{st}$). Thus, it follows:

$$Z_t = \sum_{n=0}^{\infty} L^n G_n u_t \tag{2}$$

where:

$Z_t = \begin{bmatrix} \Delta y_t \\ \Delta \pi_t \end{bmatrix}$, $L^n$ is a lag operator, $u_t = \begin{bmatrix} u_{d_t} \\ u_{s_t} \end{bmatrix}$ and $G_n$ represents the impulse-response functions associated with the endogenous variables, with the effects originating from the aggregate demand and supply shocks.

For a better representation of the model, it can be also expressed in matrix form as:

$$\begin{bmatrix} \Delta y_t \\ \Delta \pi_t \end{bmatrix} = \begin{bmatrix} G_{11}(L) & G_{12}(L) \\ G_{21}(L) & G_{22}(L) \end{bmatrix} \begin{bmatrix} u_{d_t} \\ u_{s_t} \end{bmatrix} = \sum_{k=0}^{\infty} L^k \begin{bmatrix} g_{11}(k) & g_{12}(k) \\ g_{21}(k) & g_{22}(k) \end{bmatrix} \begin{bmatrix} u_{d_t} \\ u_{s_t} \end{bmatrix} \quad (3)$$

Blanchard and Quah proposed the long-run restriction of aggregate demand influence on production, therefore the impact of demand shocks on GDP is temporary. In this respect, I applied the restriction $g_{11}(k) = 0$. Other two applied restrictions consists in the unitary variance of the shocks, requiring to match the shocks variance to 1, and in the nullity of the shocks covariance. Covariance nullity is a tool used to ensure that shocks are not intercorrelated. The covariance (cov) and variance ($s^2$) matrix of structural innovations will be expressed as follows:

$$A = \begin{bmatrix} s^2_{u_{d_t}} & cov(u_{d_t}, u_{s_t}) \\ cov(u_{d_t}, u_{s_t}) & s^2_{u_{s_t}} \end{bmatrix} = \begin{bmatrix} 1 & 0 \\ 0 & 1 \end{bmatrix} \quad (4)$$

Therefore, two of the constraints are related to the variance of structural shocks ($u_{d_t}$ and $u_{s_t}$), one restriction provides the nullity of the covariance and the long-term constraint is related to the temporary effect of the demand shocks on production.

Following the estimation of the VAR model, I applied the structural factorization method, which made it necessary to apply the long-run restriction in text form:
- @LR1(@u1) = 0 or, as a matrix
- $\begin{bmatrix} 0 & NA \\ NA & NA \end{bmatrix}$, where NA stands for coefficients ($g_{12}, g_{21}$ și $g_{22}$) that were estimated.

Firstly, the validation of the model involved the verification of the model stability by running the AR Roots Table test, which can be confirmed if the absolute values of the estimated model roots are less than 1. Finally, I used the residual diagnostic tests for the correct representation of them, as follows:
a) normal distribution of residuals through structural factorization;
b) homoskedasticity by applying the White Heteroskedasticity test;
c) the absence of residuals autocorrelation by running the Autocorrelation LM Test.

In order to confirm the aggregate demand - aggregate supply model in the PIIG countries, I have run the impulse-response function (by using the display of the cumulative effect of structural shocks) for a 10-quarter period, for each country included in the analysis. Following the identification of the effects of demand and supply positive shocks on production or inflation, it was necessary to decompose the two structural innovations included in the model in order to allow processing of the Pearson statistical correlation for the demand and supply shocks, in the analyzed countries. Considering the relation *Ae = Bu* (with matrix B being identified), according to the methodology of Blanchard and Quah, where *e* represents the residuals of the VAR model, I extracted the error terms $e_{1_t}$ and $e_{2_t}$ by using the Make residuals function, and then I calculated aggregate demand shocks and aggregate supply shocks, depending on the expressions resulting from the mentioned relation:

$$\begin{bmatrix} 1 & 0 \\ 0 & 1 \end{bmatrix} \begin{bmatrix} e_{1_t} \\ e_{2_t} \end{bmatrix} = \begin{bmatrix} b_{11} & b_{12} \\ b_{21} & b_{22} \end{bmatrix} \begin{bmatrix} u_{d_t} \\ u_{s_t} \end{bmatrix} \quad (5)$$

$$e_{1_t} = b_{11} u_{d_t} + b_{12} u_{s_t} \quad (6)$$

$$e_{2_t} = b_{21} u_{d_t} + b_{22} u_{s_t} \quad (7)$$

The Pearson statistical correlation of the structural shocks for the period 1995-2015, as well as for the two mentioned sub-periods, at 10 and 9-year intervals, was computed according to the formulas:

$$Pearson\ corr(x, y) = \frac{cov(x, y)}{s_X s_Y} \quad (8)$$

$$cov(x, y) = \frac{\sum_{i=1}^{n}(x_i - \bar{x})(y_i - \bar{y})}{n - 1} \quad (9)$$

$$s_X = \left[ \frac{\sum_{i=1}^{n}(x_i - \bar{x})}{n} \right] \quad (10)$$

where $s_X$ and $s_Y$ represents the standard deviation of the selected variable in country $X$ and $Y$.

## 4. Results and interpretations

This section examines the aggregate demand shocks and the aggregate supply shocks and their impact on the gross domestic product and inflation, by using SVAR methodology. I have applied the long-run restriction depending on the economic theory, as the impact of the aggregate demand shock on the GDP level is only temporary.

First, I have attached the results of the stationarity test for PIIG countries, both for GDP and for inflation. As Table 1 indicates, the GDP indexes for the 4 countries were not stationary at the initial level and it was required to test the stationarity for their first and second differences. The GDP series were stationary at the second difference for Greece and Ireland, while for Italy and Portugal the first difference was processed.

**Table 1**. *Stationarity*

| Country | Order of integration | ADF | Probability |
|---|---|---|---|
| Greece (GDP) | I(2) *** | -7.064368 | 0.0000 |
| Inflation | I(1) *** | -3.305331 | 0.0013 |
| Ireland (GDP) | I(2) *** | -5.147848 | 0.0000 |
| Inflation | I(1) *** | -2.948940 | 0.0037 |
| Italy (GDP) | I(1) *** | -4.583029 | 0.0000 |
| Inflation | I(0) * | -3.363538 | 0.0155 |
| Portugal (GDP) | I(1) *** | -3.173864 | 0.0019 |
| Inflation | I(0) ** | -3.682580 | 0.0299 |

*denotes the constant; ** represents the trend and constant; *** denotes the absence of trend and constant.
**Source:** Own calculations using Eviews 7.0, Eurostat (2016).

The GDP deflator in Italy and Portugal was stationary at the initial level, while the series pertaining to its evolution in Greece and Ireland involved computing the first difference, in order to obtain stationary series. The variables were stationary at the highest threshold of significance in most cases, except for the GDP deflator in Italy (1.55% probability) and Portugal (2.99% probability). In those cases, the confidence in the stationarity of the initial level of variables is greater than 95%, but less than 99%. This leads to the fact that the prices were more stable in Italy and Portugal than in Greece and Ireland.

As for choosing the optimal lag, I have used Lag Length Criteria and Lag exclusion tests, the results being displayed in Table 2. The GDP and inflation rate series are lagged by 4 quarters in Greece and Portugal, while Ireland has 5 lags and Italy - 2 lags. In the cases of Ireland and Italy, the use of Lag Length Criteria test was not sufficient in order to make a decision, and I have used the Lag exclusion test for the lag that removes heteroskedasticity from the model.

**Table 2.** *Optimal lag*

| Country | Lag Length Criteria | | | | | Lag exclusion Test (probability) |
|---|---|---|---|---|---|---|
| | Likelihood ratio | Final Prediction Error | Akaike information criterion | Schwarz information criterion | Hannan-Quinn information criterion | |
| Greece | 4 | 4 | 4 | 2 | 3 | *4 - 0.006738* |
| Ireland | 5 | 8 | 8 | 1 | 5 | *5 - 0.022051* |
| Italy | 7 | 8 | 8 | 2 | 2 | *2 - 0.012971* |
| Portugal | 4 | 4 | 4 | 1 | 4 | *4 - 0.000362* |

Lag Length Criteria test uses the selected lag by most tests. To validate it, I used Lag Exclusion Test. The chosen lag is optimal if probability is less than 5%.
**Source:** Own calculations using Eviews 7.0, Eurostat (2016).

By running the SVAR model, the resulting coefficients, ($g_{12}, g_{21}$ and $g_{22}$) proved to be accurately represented at the highest threshold of significance. I have used the coefficients of matrix B (Table 3) to decompose the residuals of the models into aggregate demand shocks and aggregate supply shocks, according to the relation *Ae = Bu*, presented in the methodology.

In order to test the stability of the model, I have attached the results of the AR Roots Graph (Figure 1). The 4 estimated models have been stable, given that the inverse of the VAR roots is less than 1 for each of them. Certain estimations based on impulse-response function cannot be validated if one of the 4 models include unit roots.

**Table 3.** *B matrix coefficients*

| Country | $b_{11}$ | $b_{12}$ | $b_{21}$ | $b_{22}$ |
|---|---|---|---|---|
| Greece | 0.001886505 | 0.011731268 | 0.009502359 | -0.003878571 |
| Ireland | 0.005303614 | 0.003396020 | 0.003951492 | -0.003881173 |
| Italy | 0.002275846 | 0.005699958 | 0.002846330 | -0.003099909 |
| Portugal | 0.002284893 | 0.006229577 | 0.004061853 | -0.000657299 |

**Source:** Own calculations using Eviews 7.0, Eurostat (2016).

**Figure 1.** *Stability of the models*

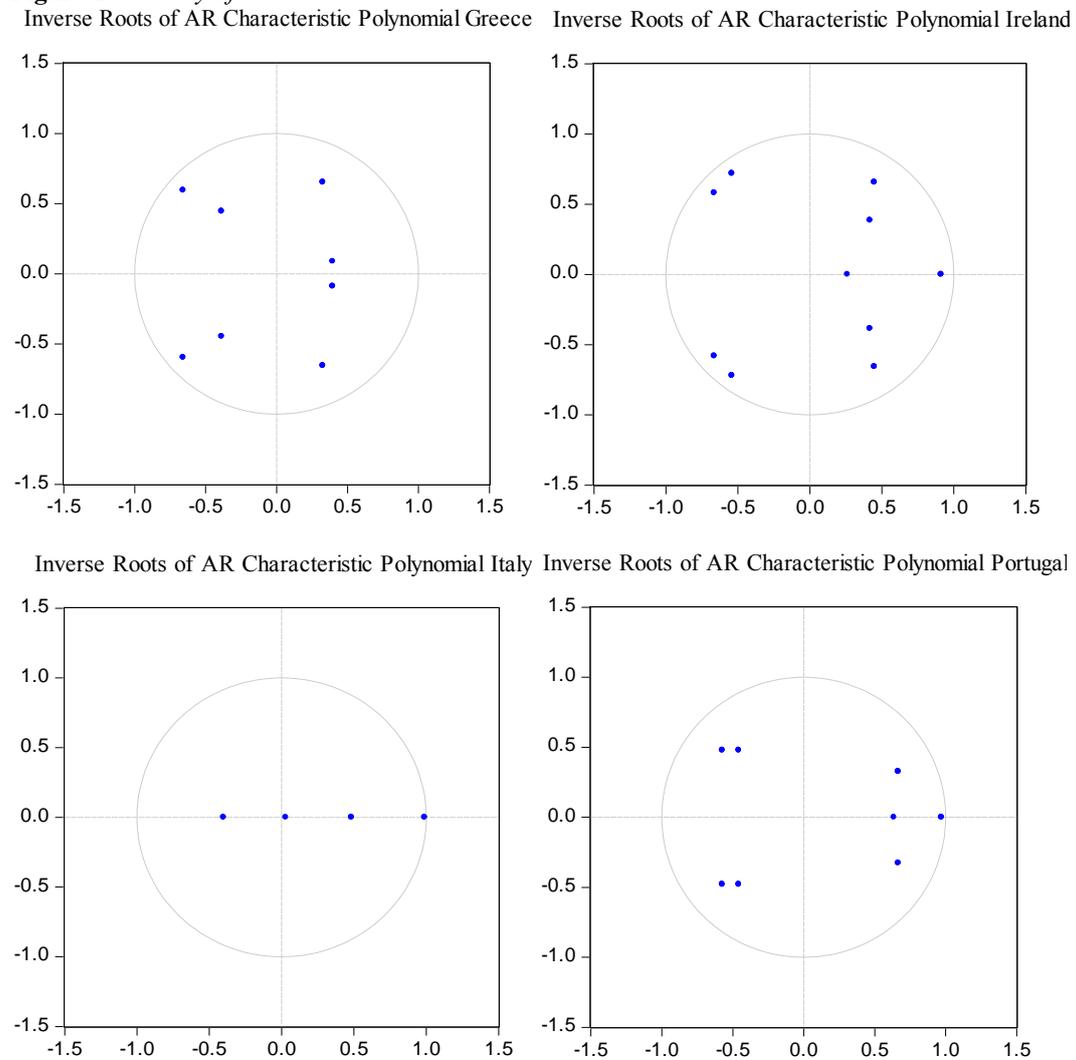

**Source:** Own calculations using Eviews 7.0, Eurostat (2016).

In Table 4, I have attached the results of residual tests. The null hypothesis of testing the normality of residuals conveys the normal distribution of error terms, which is validated when the probability of the Jarque-Berra test is greater than 5%. Jarque-Berra probability has confirmed the normal distribution of error terms in the PIIG countries and Skewness, respectively Kurtosis statistics are close to the reference values 0 and 3.

The error terms are accurately represented when the model is homoskedastic, which implies that the variance of residuals is constant. The null hypothesis of the White test consists in the homoskedasticity of the model, and the alternative one in its heteroskedasticity. The null hypothesis was validated, as the residuals were correctly represented, given that the probability of the White test exceeds the threshold of 5% in the case of the PIIG countries.

One last hypothesis to be confirmed for validating the residuals is represented by the absence of autocorrelation. For this purpose, I have performed the LM test, which sets the previously mentioned condition as a null hypothesis, and the autocorrelation assumption as an alternative one. The absence of residual autocorrelation was confirmed, given that the probability of the LM test is greater than 5%. At this point, it can be said that the estimated models have been validated, which makes it possible to analyze the impulse-response function.

**Table 4.** *Residuals diagnostic*

| Country | Normality test (structural factorization - Jarque-Berra probability) | Homoskedasticity (White Heteroskedasticity Test - probability) | Autocorrelation (LM Test - probability) |
|---|---|---|---|
| Greece | 0.2834 | 0.4505 | 0.7661 |
| Ireland | 0.2321 | 0.4071 | 0.7629 |
| Italy | 0.1432 | 0.0580 | 0.6071 |
| Portugal | 0.4638 | 0.7182 | 0.3128 |

The residuals are normally distributed if the probability of the Jarque-Berra test is greater than 5%; the model is homoskedastic if the probability of the White test is greater than 5%; the absence of autocorrelation is confirmed if the probability of the LM test is greater than 5%.
**Source:** Own calculations using Eviews 7.0, Eurostat (2016).

As Figure 2 indicates, in Greece, a positive shock in supply of 1 standard deviation point, has led to a 0.0117 units increase in the economic growth rate in the first quarter of reaction after the event of supply shock, also representing the strongest influence. The supply shock has followed a W curve, the lowest impact on GDP being manifested in the $5^{th}$ quarter after the shock (+0.0008 standard deviation points). Meanwhile, the demand shock has followed the same W curve, an 1 unit increase in the aggregate demand leading to a positive effect on the growth rate in the first 3 quarters – 0.0019, 0.0007, and 0.0014 standard deviation points. The impact was negative in the $4^{th}$ quarter (-0.0011 points), then being followed by a fluctuation between positive and negative values, from one quarter to another, until the effect of demand was neutralized. Therefore, the temporary shock has been validated in the case of aggregate demand, while the positive supply shock has set a new equilibrium level for the GDP.

In the case of Ireland, the demand shock of 1 standard deviation point (+), has resulted in an initial impact (+0.0053) on the growth rate, greater than the initial influence thereon, resulting from the supply shock (+0.0034 points), but the demand shock will however decrease in intensity. The most significant GDP change as a response to the demand shock has occurred in the $4^{th}$ quarter (+0.0058 standard deviation points) and the minimum impact from the aggregate supply increase has emerged in the $6^{th}$ quarter (+0.0024 standard deviation points). The effect of aggregate demands growth was temporary, being hardly neutralized, while increasing aggregate supply has conducted to a new equilibrium level of output (+0.0043 units).

In Italy, the initial impact of the demand shock on the GDP was 0.0023 standard deviation points, in a positive direction, and the supply shock has given an additional of 0.0057 points to the growth rate. In the first 10 quarters, after the event of supply shock, it has recorded a positive impact on the growth rate, with a maximum 0.0119 of standard deviation points in the last quarter, while the demand shock has reached its maximum contribution to the GDP growth in the $5^{th}$ and $6^{th}$ quarters (0.0039 points) after the shock.

In the $1^{st}$ quarter, Portugal has presented a positive demand shock impact of +0.0023 units on the GDP, while a positive supply shock of 1 standard deviation point has led to a 0.0062 points GDP increase. The aggregate demand has reached its maximum contribution to the GDP growth rate in the $7^{th}$ quarter (0.0057 points), while the increase of aggregate supply has resulted in a maximum +0.0106 of standard deviation point effect on the economic growth rate in the $4^{th}$ quarter.

In the case of the 4 economies, the permanent effect of supply shock on the GDP has been validated, while the demand shocks were temporary, in spite of the fact that they hardly been neutralized. Moreover, the most significant positive reaction of the gross domestic product was due to the increase in aggregate supply.

Greece is the only country in the analyzed group that surprised a cumulated impact of the aggregate demand on the economic growth rate, both positive and negative, the negative influence being caused by the persistency of inefficient budgetary policies promoted by the Hellenic economy. The Greek government has made many inefficient budgetary expenditures that have stimulated the growth of unemployment rate and underproduction, which were also accompanied by external pressures of the international creditors.

Following the analysis of individual effects (and not cumulated, as previously presented), I have found that the PIIG countries have a poor capacity to neutralize the impact of aggregate supply on the economic growth rate (long-term impact), which can be a possible effect of the rigidity of such economies. Italy needs 3 years to neutralize the influence of supply shock on the GDP, while Greece needs 10 and a half years. On the other hand, in Portugal, a supply shock on the GDP is neutralized in 5 years, while Ireland requires a much longer adjustment period (8 years and 9 months).

**Figure 2.** *Response of gross domestic product to structural innovations*

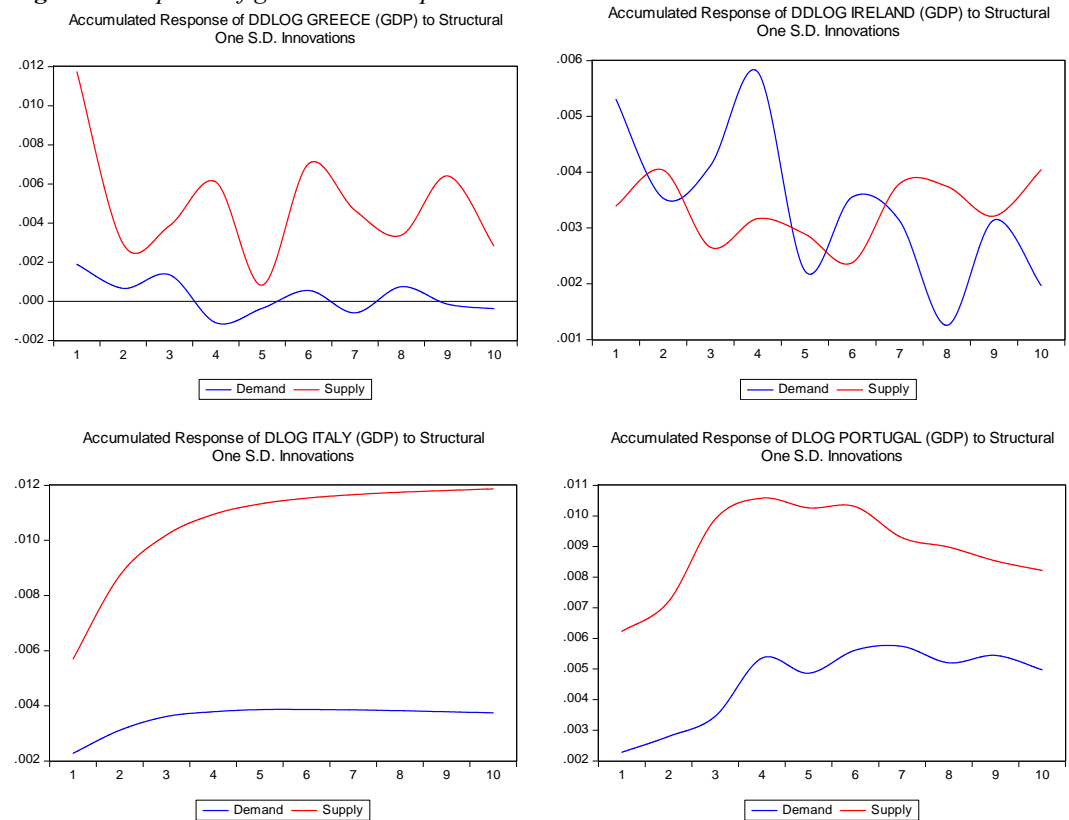

**Source:** Own calculations using Eviews 7.0, Eurostat (2016).

The cumulated impact of aggregate demand and supply on the inflation has been analyzed based on the graphs processed in Eviews and inserted in Figure 3. In Greece, a 1 standard deviation point of positive demand shock had an initial impact of +0.0095 points on the growth rate of GDP deflator, and a similar-intensity positive supply shock has generated 0.0039 of units decrease in the inflation rate in the first quarter after the shock. In the first 10 quarters, the maximum impact of the demand shock on the inflation rate has occurred in 8th and 10th quarters (0.0122 standard deviation points). Regarding an 1 unit positive supply shock, the strongest impact on the inflation rate has emerged in 4th and 7th quarters (-0.0073 points).

In Ireland, the initial impact of the demand shock on inflation rate was +0.004 of standard deviation points, reaching a maximum +0.023 of points in the 10th quarter (in the first 10 quarters). An 1 unit growth in aggregate supply has confirmed the aggregate demand – aggregate supply model due to the negative response of inflation rate in the 1st quarter, of 0.004 standard deviation points. The most drastic decrease in the inflation rate in the first 10 quarters has occurred in the last specified quarter (0.012 points).

In Italy, the effects were similar to those in Ireland, with an initial impact of +0.003 and -0.003 units for a 1 standard deviation point increase in aggregate demand and aggregate supply. Italy has felt the strongest impact on reducing prices, due to an increase in aggregate supply.

Finally, Portugal has presented the lowest impact of the aggregate supply (+1 standard deviation point) on the inflation rate (-0.001 units). On the demand side, the impact on the price growth rate was 0.004 units in the 1$^{st}$ quarter after the shock.

**Figure 3.** *Response of inflation to structural innovations*

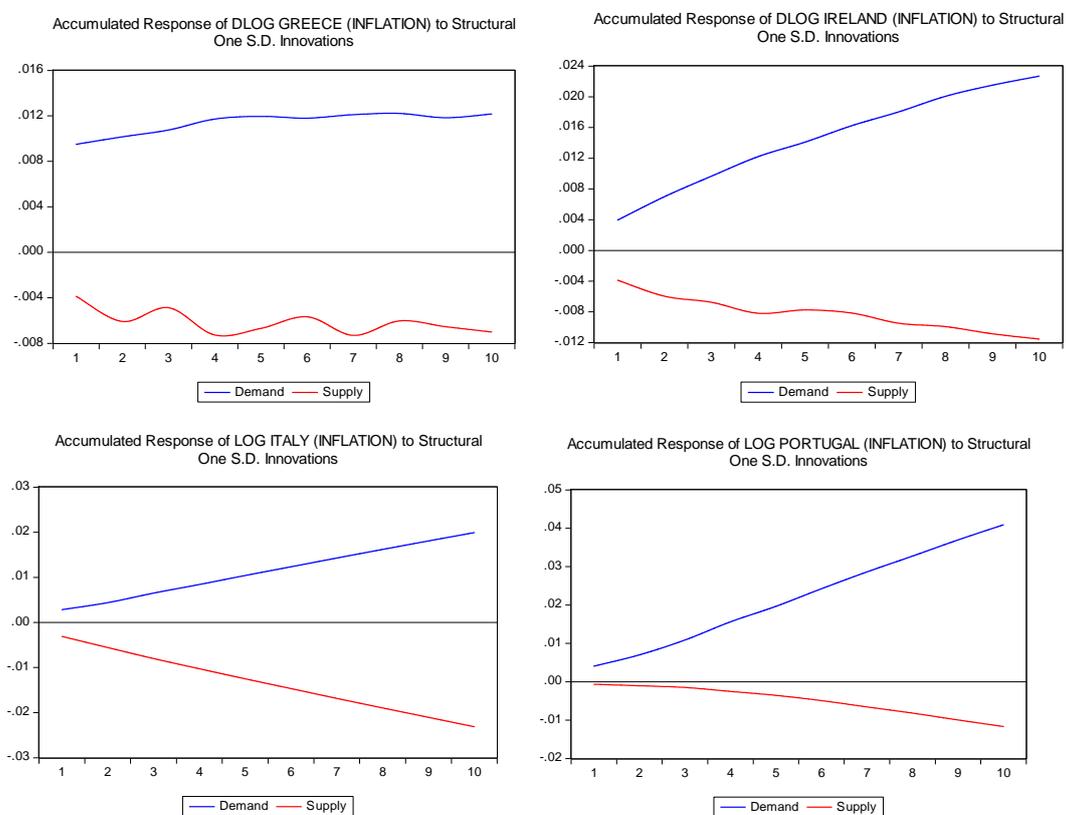

**Source:** Own calculations using Eviews 7.0, Eurostat (2016).

The impact of aggregate demand and aggregate supply on the inflation rate is hardly neutralized in the PIIG countries, which points to greater costs for them for being members of the EMU. In other words, the analyzed countries have a low economy adjustment capacity, and, by adopting the common European currency, they turned into losers of the monetary integration process, as they have lost an important part of monetary policy sovereignty.

The aggregate demand - aggregate supply model was validated in the 4 countries, regarding the response of both prices and gross domestic product to the aggregate demand and supply shocks.

*Annex 1* includes the structural innovations for each estimated model. In order to provide a broader view in the volatility of pre-crisis and post-crisis aggregate demand and aggregate supply shocks, I have calculated the standard deviation for each structural shock. In the pre-crisis period, Greece has presented a 0.9978 of standard deviation of aggregate demand shock, which has exceeded the standard deviation of aggregate supply shock (0.9141). In the post-crisis period, the supply shocks were more volatile, with a standard deviation increasing to 0.9758, while the standard deviation of supply shocks decreased to 0.8012 points, an inversion that was also seen in the case of Italy.

Ireland and Portugal have been in the opposite position, as the demand shocks in these countries were more volatile than these of supply shocks, both in pre-crisis, and in post-crisis periods. In Ireland, the standard deviation of demand shocks increased from 0.7854 (1997-2006) to 1.0380 (2007-2015), while the standard deviation of supply shocks increased from 0.9211 to 0.9387 points.

PIIG economies were strongly affected in the $1^{st}$ quarter of 2009, as a result of the economic crisis, by negative shocks both on the demand and on the supply side.

Greece recorded negative shocks in both demand and supply following the adoption of the Euro currency, as well as in the $3^{rd}$ quarter of 2010 and in the $4^{th}$ quarter of 2014. Following the collapse of the real estate market, Ireland witnessed negative shocks in aggregate demand starting with the $2^{nd}$ quarter of 2007, which were hardly absorbed. The aggregate supply was revived by the reduction of the production factor costs. In the $1^{st}$ quarter of 2005, Italy's economy stagnated due to the shrinking of investments and persistence of a pessimistic economic sentiment, caused by the unfavorable position of public finances. Throughout the year, the aggregate demand recovered due to the spendings made by the Italian government. In the same period, Portugal had a difficult position, with the aggregate supply significantly dropping in the $3^{rd}$ quarter of 2005, while an improvement occurred in terms of aggregate demand. This evolution was mainly caused by the recovery of the Portuguese exports and a significant decrease of imports.

In order to analyze the synchronization of demand and supply shocks, I have attached the results of the Pearson correlation in the period 1997-2015 and by sub-periods (1997-2006 and 2007-2015) in Table 5. The processing of the bilateral correlations has resulted in 6 correlations both of demand and supply side.

The highest level of aggregate supply shocks correlation has been recorded between Portugal and Italy, while the country pair Italy-Greece manifested the most significant extent of demand shock synchronization.

During the period 1997-2015, the correlation matrix has been balanced, with 3 correlations presenting higher levels in the supply side, and 3 favoring the synchronism of demand shocks. The balance did not change in the sub-periods mentioned-above, except for the country pairs Ireland-Greece and Italy-Ireland. Basically, in 1997-2006 period,

Ireland presented a higher extent of synchronization of the demand shocks along Greece, and, following the emergence of the financial crisis, the correlation of supply shocks became greater than the synchronization of demand shocks. On the other hand, in the period 1997-2006, the correlation of the supply shocks between Italy and Ireland has been greater than the correlation of demand shocks. The trend has been inverted in the period 2007-2015. However, the most significant repositioning of the structural shocks correlation has been seen on the evolution of synchronism of demand shocks (from a 13.08% divergence in the sub-period 1997-2006 to a 25.02% convergence in the sub-period 2007-2013) and supply shocks (from a 10.94% divergence in the sub-period 1997-2006 to a 30.43% correlation in the sub-period 2007-2013) between Portugal and Greece. Moreover, the financial crisis had the effect of increasing the structural innovations correlations for most pairs of countries included in the analysis.

Overall, it can be said that the country pairs Ireland-Greece, Greece-Portugal and Ireland-Portugal are more correlated in terms of aggregate demand shocks, which means a greater similarity extent between the adopted economic policies and the macroeconomic situation. Regarding the synchronism of the supply shocks, the country pairs Greece-Italy, Italy-Ireland and Italy-Portugal are better positioned, which can be explained by greater structural similarities between them and by a greater convergence of the business cycles.

The research conducted in this field has proven that the economies of the European Union presents several intercorrelated supply shocks, with the financial crisis manifesting an impulse in this respect. In that case, PIIG economies are correlated both in demand and in supply, with the balance of bilateral correlations of structural shocks, being equally distributed. This situation may be blamed on the high level of public debt and the many recommendations submitted by the "Troika" primarily to Greece, but also to Portugal, Italy and Ireland. The high similarity between them in terms of disastrous public finances situation has led to more synchronized economic policies.

**Table 5.** *Demand shocks correlation and supply shocks correlation*

| Country | Greece | Ireland | Italy | Portugal |
|---|---|---|---|---|
| *Correlation 1997-2015* | | | | |
| Greece | 100.00% | 07.54% | 23.90% | 13.67% |
| Ireland | 07.46% | 100.00% | 20.13% | 16.43% |
| Italy | 38.96% | 21.38% | 100.00% | 17.14% |
| Portugal | 04.76% | -01.22% | 44.40% | 100.00% |
| *Correlation 1997-2006* | | | | |
| Greece | 100.00% | -00.39% | 18.19% | -10.94% |
| Ireland | 15.73% | 100.00% | 25.74% | 10.54% |
| Italy | 38.07% | 16.04% | 100.00% | 20.24% |
| Portugal | -13.08% | -07.13% | 41.83% | 100.00% |
| *Correlation 2007-2015* | | | | |
| Greece | 100.00% | 09.63% | 20.34% | 30.43% |
| Ireland | 00.38% | 100.00% | 13.37% | 18.68% |
| Italy | 39.59% | 26.14% | 100.00% | 16.00% |
| Portugal | 25.02% | 05.47% | 48.82% | 100.00% |

☐ Demand shocks correlation   ☐ Supply shocks correlation
**Source:** Own calculations using Eviews 7.0, Eurostat (2016).

## 5. Conclusions

The situation of Greece in the recent years, as well as the high levels of public debt in Ireland, Italy and Portugal has laid pressure on the foundations of the Economic and Monetary Union.

This study validates the aggregate demand – aggregate supply model, also demonstrating the limited capacity of the mentioned economies to coordinate their economic cycles, given the sovereignty transfer to ECB, by the common monetary policy. The poor capacity to neutralize demand and supply shocks is also felt in the case of the impact on the economic growth rate, not only in the case of price influencing. The long-term neutralization of shocks in PIIG countries reinforces the idea that Greece, Ireland, Italy and Portugal recorded considerable losses relative to the obtained gains following their adhesion to the Euro zone.

After the effects of the economic crisis have been experienced in Italy and Greece, the aggregate supply shocks have become more volatile than the aggregate demand shocks, this pair of country recording a higher extent of synchronization of aggregate supply shocks. Compared to the Euro zone, which presents a higher correlation of supply shocks between the euro area members, PIIG countries are characterized by an equilibrium in the synchronism of aggregate demand shocks and aggregate supply shocks. This was caused by the macroeconomic similarities, regarding the situation of public debts and budget deficits, which directed those countries to adopt similar economic measures.

It is clear that the subjects of the study behave differently from the core of the Euro zone and, in the absence of a fiscal union, the idea to reduce the asymmetry between member states remain just an utopia. A fiscal union would play an important role in the federalist future of the EU, which could also solve the issue of fragmentation among it, known as one of its greatest challenges.

## References

Bayoumi, T., 1992. The effect of the ERM on Participating Economies, *IMF Staff Papers*, Vol. 39, No. 2, pp. 330-356.

Bayoumi, T., Eichengreen, B., 1992. Shocking Aspects of European Monetary Integration, *NBER Working Paper*, No. 3949.

Bergman, M., 2005. Do-lung run restrictions identify supply and demand disturbances?, mimeo, October.

Blanchard, J., Quah, D., 1989, The Dynamic Effects of Aggregate Demand and Supply Disturbances, *The American Economic Review*, Vol. 79, No. 4, pp. 655-673.

Cooley, T., Dwyer, M., 1998. Business Cycle Analysis Without Much Theory: A Look at Structural VARs, *Journal of Econometrics*, Vol. 83, No. 1-2, pp. 57-88.

Dinu, M., Marinaș, M., 2006. Economic Transformation of European Union in the Context of Kondratieff Cycles, *Theoretical and Applied Economics*, No. 8(503), pp. 29-36.

Fidrmuc, J., Korhonen, I., 2003. Similarity of supply and demand shocks between the euro area and the CEEC, *Economic Systems*, Vol. 27, No. 3, pp. 313-334.

Frenkel, M., Nickel, K., 2005. How Symmetric are the Shocks and the Shock Adjustment Dynamics between the Euro Area and Central and Eastern European Countries?, *Journal of Common Market Studies*, Vol. 43, No. 1, pp. 53-74.

Fry, R., Pagan, A., 2011. Sign Restrictions in Structural Vector Autoregressions: A Critical Review, *Journal of Economic Literature*, Vol. 49, No. 4, pp. 938-960.

Gottschalk, J., 2001. An Introduction into the SVAR Methodology: Identification, Interpretation and Limitations of SVAR models, *Kiel Working Paper*, No. 1072.

Marinaș, M., 2012. Testing the Asymmetry of Shocks with Euro Area, *Theoretical and Applied Economics*, Vol. XIX, No. 1(566), pp. 5-20.

Marinaș, M., 2013. *Convergența reală și sincronizarea ciclurilor de afaceri cu zona euro*, Ed. ASE, București.

Marinaș, M., Socol, A., Socol, C., 2010. Fiscal Sustainability and Social Cohesion. Common and Specific in EU Sub-models, *Theoretical and Applied Economics*, Vol. XVII, No. 3(544), pp. 43-62.

Socol, A., 2011. Costs of Adopting a Common European Currency. Analysis in Terms of the Optimum Currency Areas Theory, *Theoretical and Applied Economics*, Vol. XVIII, No. 2(555), pp. 89-100.

Eurostat, 2016. Eurostat Database, [online], available at: <www.ec.europa.eu/eurostat/data/database/> [Accessed 15th April 2016]

## Annex 1. Aggregate demand and supply shocks

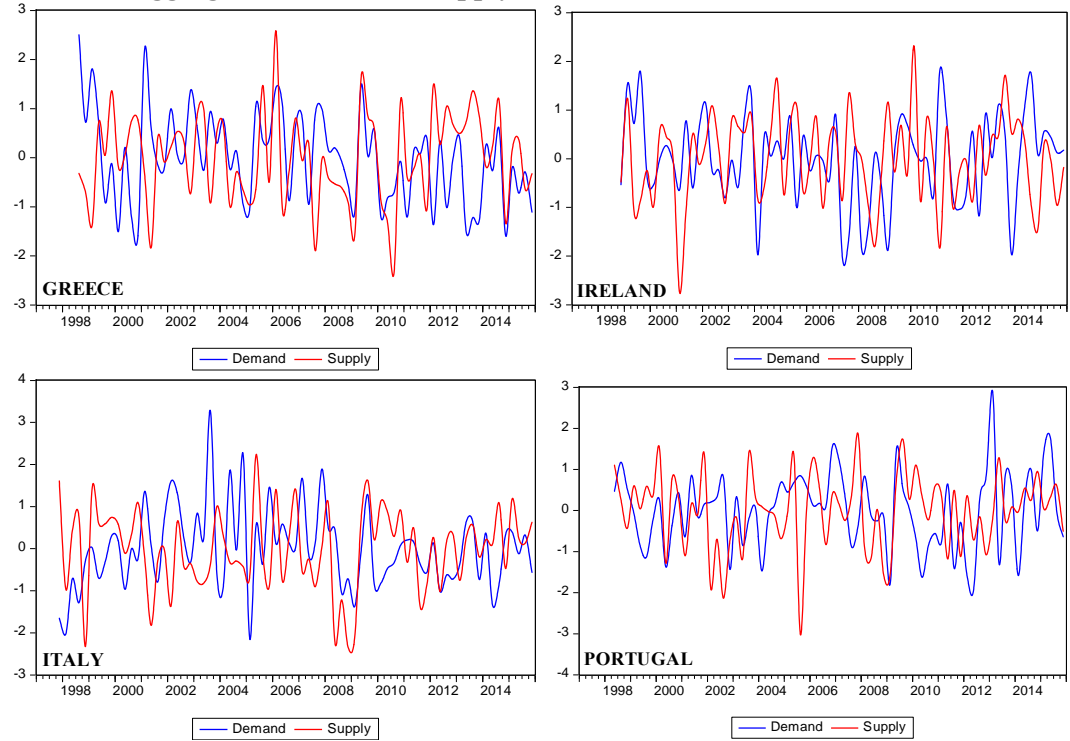

**Source:** Own calculations using Eviews 7.0, Eurostat (2016).